\documentclass[aps,prb,twocolumn]{revtex4}

\usepackage{graphics,epsfig} 

\begin{document}

\title{Einstein's Boxes}
\author{Travis Norsen}
\email{norsen@marlboro.edu}
\affiliation{Marlboro College, Marlboro, Vermont 05344}

\date{\today}

\begin{abstract}
At the 1927 Solvay conference, Albert Einstein presented a thought 
experiment intended to demonstrate the incompleteness of the 
quantum mechanical description of reality. In the following
years, the experiment was modified by Einstein, de Broglie, and 
several other commentators into a simple scenario involving the 
splitting in half of the wave function of a single particle in a 
box. This paper collects together several formulations of this 
thought experiment from the 
literature, analyzes and assesses it from the point of view of the
Einstein-Bohr debates, the EPR dilemma, and Bell's theorem, and 
argues for ``Einstein's Boxes'' taking its rightful place 
alongside similar but historically better known quantum
mechanical thought experiments such as EPR and Schr\"odinger's
Cat.
\end{abstract}

\maketitle

\section{Introduction}

It is well known that several of quantum theory's founders
were dissatisfied with the theory as interpreted by Niels Bohr and
other members of the Copenhagen school. Before about 1928, for
example, Louis de Broglie advocated what is now called a
hidden variable theory: a pilot-wave version of quantum mechanics
in which particles follow continuous trajectories, guided by a
quantum wave.\cite{jammer} David Bohm's\cite{bohm} rediscovery and
completion of the pilot-wave theory in 1952 led de Broglie back to
these ideas; Bohm's theory has continued to inspire interest in
alternatives to the Copenhagen interpretation.\cite{bell} Erwin
Schr\"odinger was likewise doubtful that the quantum wave function
could alone constitute a complete description of physical reality.
His famous ``cat'' thought experiment\cite{schrod} was intended to
demonstrate quantum theory's incompleteness by magnifying the
allegedly real quantum indefiniteness up to the macroscopic level
where it would directly conflict with experience.

By far the most important critic of quantum theory, however, was
Albert Einstein.\cite{oppose} Several of his early thought
experiments attacking the completeness doctrine are memorably
recounted in Bohr's reminiscence.\cite{bohr}  Einstein's most
important argument against quantum theory's completeness is the
paper (EPR) he co-authored with Boris Podolsky and Nathan Rosen in
1935.\cite{epr}

The purpose of this paper is to resurrect another thought 
experiment which, like EPR and Schr\"odinger's cat, is intended to
argue against the orthodox doctrine of quantum completeness. This 
thought experiment -- ``Einstein Boxes''\cite{fine} -- is due
originally to Einstein, although it has also been discussed and 
reformulated by de Broglie, Schr\"odinger, Heisenberg, and others. 

Given its unique simplicity, clarity, and elegance, the relative
obscurity of this thought experiment is unjustified. Although
generally aiming to establish the same dilemma as that posed in
the EPR paper,\cite{epr} Einstein's Boxes establishes this
conclusion with a more straightforward logical argument. Our hope
is that this paper will aid in re-injecting this old thought
experiment into the ongoing discussions of the significance and
implications of EPR, the completeness doctrine, and alternatives
to Copenhagen such as Bohm's non-local hidden variable theory.
Because of its remarkable simplicity, the Einstein Boxes thought
experiment also is well-suited as an introduction to these topics
for students and other interested non-experts.

The paper is organized as follows. In Sec.~\ref{sec:eprdebroglie}
we introduce Einstein's Boxes by quoting a detailed description
due to de Broglie. We compare it to the EPR argument and discuss
how it fares in eluding Bohr's rebuttal of EPR. In
Sec.~\ref{sec:einstein} we present and discuss Einstein's
original version of the thought experiment as well as some
of his related comments regarding the boxes and their
connection to EPR. In Sec.~\ref{sec:bell} we present a new,
Bell-inspired formulation of Einstein's Boxes and relate the ideas
to Bell's celebrated theorem about hidden variable theories. 
Section~\ref{sec:locality} begins with some comments of
Heisenberg on the thought experiment, considers an experimentally
realized version of it, and assesses some
of Heisenberg's statements on non-locality. In
Sec.~\ref{sec:discussion} we conclude with a general
discussion.

\section{EPR and de Broglie's Version of the Boxes}
\label{sec:eprdebroglie}

\begin{figure}[t]
\begin{center}
\epsfig{file=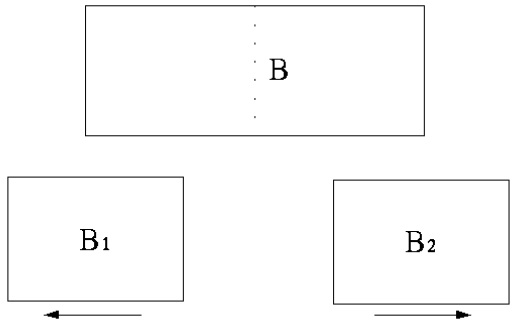, width=.45\textwidth}
\end{center}
\caption{
\label{fig:splitboxes}
A single particle is confined to the box $B$, into which a barrier
is inserted thus splitting the box and the particle's wave function
in two. The two half boxes ($B_1$ and $B_2$) are then separated, 
at which point the boxes may be opened and their contents examined.}
\end{figure}

In a 1964 book de Broglie gave a detailed statement of the 
Einstein's Boxes thought experiment.\cite{hardy}
\begin{quote}
``Suppose a particle is enclosed in a box $B$ with impermeable
walls. The associated wave $\Psi$ is confined to the box and
cannot leave it. The usual interpretation asserts that the
particle is ``potentially'' present in the whole of the box $B$,
with a probability $|\Psi|^2$ at each point. Let us suppose that
by some process or other, for example, by inserting a partition
into the box, the box $B$ is divided into two separate parts $B_1$
and $B_2$ and that $B_1$ and $B_2$ are then transported to two
very distant places, for example to Paris and Tokyo. 
[See Fig.~\ref{fig:splitboxes}.]  The particle,
which has not yet appeared, thus remains potentially present in
the assembly of the two boxes and its wave function $\Psi$ 
consists of two parts, one of which, $\Psi_1$, is located in $B_1$
and the other, $\Psi_2$, in $B_2$. The wave function is thus of
the form $\Psi = c_1 \Psi_1 + c_2 \Psi_2$, 
where $|c_1|^2 + |c_2|^2 = 1$.

``The probability laws of wave mechanics now tell us that if an 
experiment is carried out in box $B_1$ in Paris, which will enable the 
presence of the particle to be revealed in this box, the probability of 
this experiment giving a positive result is $|c_1|^2$, whilst the 
probability of it giving a negative result is $|c_2|^2$. According to 
the usual interpretation, this would have the following significance: 
because the particle is present in the assembly of the two boxes prior to 
the observable localization, it would be immediately localized in box 
$B_1$ in the case of a positive result in Paris. This does not seem to 
me to be acceptable. The only reasonable interpretation appears to me 
to be that prior to the observable localization in $B_1$, we know that 
the particle was in one of the two boxes $B_1$ and $B_2$, but we do not 
know in which one, and the probabilities considered in the usual wave 
mechanics are the consequence of this partial ignorance. If we show 
that the particle is in box $B_1$, it implies simply that it was 
already there prior to localization. Thus, we now return to the clear 
classical concept of probability, which springs from our partial 
ignorance of the true situation. But, if this point of view is 
accepted, the description of the particle given by the customary wave 
function $\Psi$, though leading to a perfectly \emph{exact} description 
of probabilities, does not give us a \emph{complete} description of the 
physical reality, because the particle must have been localized prior to 
the observation which revealed it, and the wave function $\Psi$ gives 
no information about this.

``We might note here how the usual interpretation leads to a paradox in 
the case of experiments with a negative result. Suppose that the 
particle is charged, and that in the box $B_2$ in Tokyo a device has 
been installed which enables the whole of the charged particle located 
in the box to be drained off and in so doing to establish an observable 
localization. Now, if nothing is observed, this negative result will 
signify that the particle is not in box $B_2$ and it is thus in box 
$B_1$ in Paris. But this can reasonably signify only one thing: the 
particle was already in Paris in box $B_1$ prior to the drainage 
experiment made in Tokyo in box $B_2$. Every other interpretation is 
absurd. How can we imagine that the simple fact of having observed 
\emph{nothing} in Tokyo has been able to promote the localization of 
the particle at a distance of many thousands of miles
away?''\cite{db} 
\end{quote}

Although de Broglie did not make his reasoning explicit, he
gives some hints that allow us to plausibly reconstruct the
intended logical structure of this argument for incompleteness.
The rhetorical question at the end implies that de Broglie thinks
it is impossible (``absurd'') for the negative result of a
measurement in Tokyo to have a decisive causal effect on the
contents of the box in Paris. Likewise for a positive result: ``If
we show that the particle is in box $B_1$, it implies simply that
it was already there prior to localization,'' and therefore, by
implication, that it was already \emph{not} in box $B_2$ before the 
observation. Generalizing, the
measurement at $B_1$ cannot have affected the contents of the box
$B_2$, and therefore the contents of $B_2$ (which are \emph{known}
after inspection of $B_1$) must have been already 
\emph{definite} before that inspection. 

According to at least one commonly held version of the orthodox 
interpretation, however, it is the very act of observation that 
disturbs the physical system in question and brings about a
definite result. Why should de Broglie think that the act of
opening $B_1$ and examining its contents can't affect the contents 
of $B_2$? Because $B_2$ is \emph{spatially distant}. Its contents could 
not be affected by any reasonable physical mechanism by something done 
to the well-separated $B_1$. That is evidently the point of taking one 
box to Paris and the other to Tokyo before opening them.

There is thus a locality or separability assumption at work here. 
Given that $B_2$ is physically separated from $B_1$ by a great 
distance, nothing we do to $B_1$ can affect the contents of $B_2$. 
(Indeed, we could consider the contents of $B_2$ at space-time 
coordinates that lie outside the light cone of the observation event at 
$B_1$. Then any ``disturbance'' by the measurement at $B_1$ on $B_2$'s 
contents would evidently violate relativity's prohibition on superluminal 
causation.) So when we open $B_1$ and determine its contents, we learn 
immediately whether or not $B_2$ contains the particle. And because $B_2$'s 
contents cannot have been in any way affected by the opening of $B_1$,
$B_2$ must have either contained or not contained the particle all along. 
And, finally, because ``the wave function $\Psi$ gives no
information about'' the particle's actual pre-measurement location
(in particular, $\Psi$ attributes no definite particle content to
$B_2$), that description must have been \emph{incomplete}.

De Broglie seems to have had this reasoning in mind.  Notably, it is 
structurally identical to the argument of the more famous EPR 
paper.\cite{epr}  EPR begins with a 2-particle wave function 
\begin{equation}
\psi(x_1,x_2)=\delta(x_1-x_2-a) = \!\int dk \, e^{ik(x_1-x_2-a)},
\end{equation}
which is a simultaneous eigenstate of the relative position operator 
$(\hat{x}_1 - \hat{x}_2)$ (with eigenvalue $a$) and of the total momentum
operator $(\hat{p}_1 + \hat{p}_2)$ (with eigenvalue $0$). The authors
also assume the two particles to be sufficiently well-separated
that any measurement performed on particle 1 can have no physical
effect on particle 2. They then argue as follows: if the
experimenter chooses to measure the position of particle 1, he
will immediately be able to infer the position of particle 2;
should he choose instead to measure the momentum of particle 1,
he may immediately infer the momentum of particle 2. Therefore,
because nothing the experimenter does in the region near particle
1 can affect particle 2 in any way, both the position and
momentum of particle 2 must already have had definite values
prior to any measurement. Both quantities are ``elements of
reality'' according to the famous EPR criterion of reality: 
``If, without in any way 
disturbing a system, we can predict with certainty (that is, with
probability equal to unity) the value of a physical quantity,
then there exists an element of physical reality corresponding to
this physical quantity.'' \cite{epr}

Having also defined a necessary condition for the completeness
of a theory (``every element of the physical reality must have a
counterpart in the physical theory''\cite{epr}), EPR concluded that 
the quantum mechanical description is incomplete because it
doesn't permit simultaneous attribution of definite position and 
momentum values to particle 2. 

Niels Bohr is widely believed to have refuted the EPR argument by 
pointing out an ``essential ambiguity'' in EPR's criterion of
reality.\cite{bohr2}  The alleged ambiguity was contained in EPR's 
phrase ``without in any way disturbing a system.'' In his response, 
Bohr appears to concede EPR's locality/separability principle when 
he writes that ``there is in a case 
like that just considered no question of a mechanical disturbance
of the system under investigation during the last critical stage
of the measuring procedure.''
However: ``even at this stage there is essentially the question of
an influence on the very conditions which define the 
possible types of predictions regarding the future behavior of the 
system. Because these conditions constitute an inherent element of
the description of any phenomenon to which the term `physical
reality' can be properly attached, we see that the argumentation
of the mentioned authors does not justify their conclusion that
quantum-mechanical description is essentially incomplete.'' \cite{bohr2}

Bohr seems to have meant\cite{jam2} that although we do not
``mechanically'' disturb the second particle by measuring the
position or momentum of particle 1, we do enact a decisive choice
when we decide \emph{which} quantity to measure. For whichever
quantity we decide to actually measure, we forgo the possibility
of subsequently learning the value of the complementary variable
in the distant particle. Bohr discusses this decisive choice in
terms of the need for a ``\emph{discrimination between different
experimental procedures which allow of the unambiguous use of
complementary classical concepts}.''\cite{bohr2}  Bohr is evidently 
referring to the mutually exclusive experimental arrangements
needed to measure the position and momentum of particle 1. 

In short, although we can choose to learn about the position of 
particle 2 or the momentum of particle 2, we cannot with any 
single experimental arrangement learn about both the position 
and the momentum. And, says Bohr, this fact demonstrates the 
invalidity of EPR's simultaneous use of the two concepts in regard 
to particle 2: the experimental arrangement which warrants use of 
one concept makes the other simply inapplicable. Bohr believed that 
this ``complementarity'' perspective allows one to consistently 
maintain that the ``quantum-mechanical description of physical 
phenomena ... fulfill[s] ... all rational demands of completeness.'' 
\cite{bohr2}  That is, it eludes the EPR argument from locality to 
incompleteness.

Although widely accepted in the physics community, the adequacy of Bohr's 
reply to EPR has been questioned many times by those who have carefully 
scrutinized the issue. \cite{bell2}  We will not fully engage this 
long-standing dispute, but we can shed some fresh light on the
debate over the validity of EPR's conclusion by returning to Einstein's
Boxes and seeing how this simpler thought experiment fares in
escaping what appears to have been the main thrust of Bohr's
response to EPR.

It seems reasonable to infer from de Broglie's formulation of the
Boxes argument (quoted above) that de Broglie meant to assert the 
existence of a pre-measurement ``element of reality''
corresponding to the contents of (at least) the distant box,
$B_2$, and that he would argue for its existence on the
basis of the EPR criterion of reality: ``without in any way
disturbing'' box $B_2$, we can determine whether it contains
a complete particle or nothing. The presence or non-presence
of the particle in $B_2$ is therefore an element of reality even
before the box $B_1$ is opened and its contents examined. Finally,
because the initial quantum mechanical wave function $\Psi$
attributed neither (definite) state to box $B_2$, there is at least 
one element of reality which fails to ``have a counterpart in the 
physical theory.'' And therefore quantum theory is incomplete by 
precisely the EPR completeness criterion.

The logic of this elaboration of de Broglie's argument exactly
matches the logic of the EPR paper. But because the Boxes thought
experiment in no way relies on a choice between two complementary
quantities to measure,\cite{maudlin} it seems immune from Bohr's
criticism that there is a kind of ``semantic disturbance''\cite{fine2}
effected by this choice. There is no question with the Boxes
experiment about which quantity will be measured. We are going to
open $B_1$ and look for the particle there, period. So there
appears to be no room for ``\emph{an influence on the very conditions
which define the possible types of predictions regarding the
future behavior of the system}.'' \cite{bohr2} Given the 
assumption of locality/separability (with which Bohr appears to 
agree given his denial of a ``mechanical disturbance''), the 
Einstein's Boxes thought experiment establishes the conclusion of
incompleteness  (that is, it establishes the dilemma between
locality and completeness) in a straightforward way that seems
immune to Bohr's criticism of EPR.

\section{Einstein and the Boxes}
\label{sec:einstein}

\begin{figure}[t]
\begin{center}
\epsfig{file=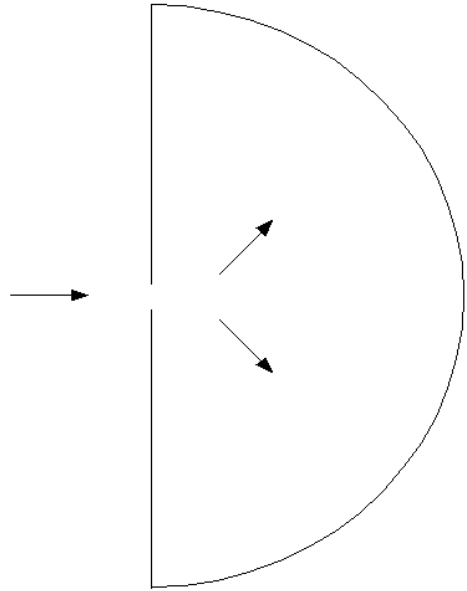, width=.4\textwidth}
\end{center}
\caption{
\label{fig:einscreen}
Einstein's original version of the boxes thought experiment: 
particles incident from the left penetrate a narrow aperture, where 
they diffract, resulting in essentially spherical Schr\"odinger waves
propagating toward the hemispherical detection screen.}
\end{figure}

Although it is not specifically formulated in terms of a particle
being split between a pair of boxes, Einstein presented a
thought experiment at the 1927 Solvay Conference\cite{jam3,ballentine} 
on which de Broglie's later formulation was based. The experiment 
involves a particle incident on a diaphragm with a single aperture,
 behind which lies a large, hemispherical detection screen.  (See
Fig.~\ref{fig:einscreen}.) With a sufficiently narrow aperture, the
incident electrons will diffract, resulting in essentially
spherical Schr\"odinger waves propagating toward the screen. If a
single electron is sent through the aperture, the electron will
eventually be detected at some distinct point on the screen.
Einstein's comments focus on the apparent conflict between the 
spreading spherical wave and the distinct point of eventual detection.

Einstein described two possible interpretations of this scenario.
According to Interpretation 1,
``The de Broglie-Schr\"odinger waves do not correspond to a single
electron, but to a cloud of electrons extended in space. The
theory does not give any information about the individual
processes, but only about the ensemble of an infinity of
elementary processes.'' In Interpretation 2,
``The theory has the pretension to be a complete theory of
individual processes.'' \cite{ball2}

Implicit in Interpretation 1 is the idea that each 
particle follows some definite trajectory from the slit to the
screen; the spreading of the wave merely indicates
ignorance about the exact identity of this trajectory for any 
particular particle. The wave represents a kind of ensemble
average over a large number of such individual trajectories. The
standard Born rule can then be understood as follows:
$|\Psi|^2$ gives the probability that any particular electron in this 
(real or imagined) ensemble in fact has (just prior to detection)
a given location. Thus, according to Interpretation 1, the
description in terms of wave and Born rule
probabilities is incomplete because it fails to describe
the actual path taken by each individual particle between the slit
and the screen.

Interpretation 2, on the other hand, denies that each individual electron
has a definite trajectory. Instead, each electron is assumed to be 
spread out in a way that is accurately described by the spherical 
wave. This interpretation was probably meant to capture the view
held by Schr\"odinger at the time,\cite{jam4}
according to which $|\Psi|^2$ represented an actual charge density for 
an individual electron. It also is a fair description of the views
of Jordan and (the later) 
Heisenberg who held that, prior to the moment of detection, 
there existed a spatially dispersed potentiality for particle localization, 
these potentialities being triggered into realities by the eventual 
interaction with a measurement device.\cite{jam5}

As Einstein went on to point out, however, the particle is in fact always
found at a definite point on the screen. Thus, at the moment it is 
detected there, the Schr\"odinger wave (representing charge
density or some kind of irreducible potentiality for particle
localization) must suddenly collapse to zero at all
other points on the screen. And, as long as we hold (with
Interpretation 2) that the initially spherical wave accurately and
completely captures the physical reality of the propagating
particle, this collapse evidently involves a kind of 
action-at-a-distance. 

Max Jammer summarizes Einstein's objection to Interpretation 2 as
follows:  ``If $|\psi|^2$ is interpreted [this way], then, 
as long as no localization has been effected, the particle must be 
considered as potentially present with almost constant probability over 
the whole area of the screen; however, as soon as it is localized, a 
peculiar action-at-a-distance must be assumed to take place which 
prevents the continuously distributed wave in space from producing an 
effect at \emph{two} places on the screen.'' \cite{jam6} 
This possibility will be 
further explored in Sec.~\ref{sec:bell}.

Einstein continued\cite{jam7}:
``It seems to me that this difficulty cannot be overcome 
unless the description of the process in terms of the Schr\"odinger wave 
is supplemented by some detailed specification of the localization of 
the particle during its propagation. I think M. de Broglie [who, 
recall, was toying with the pilot-wave theory at this time] is 
right in searching in this direction. If one works only with 
Schr\"odinger waves, the Interpretation 2 of $|\psi|^2$, I think, 
contradicts the postulate of relativity.''

The relevant aspect of the ``postulate of relativity'' is of course 
precisely the locality/separability assumption mentioned previously: 
there can be no causal dependence between two space-like separated 
events. Relativity thus prohibits the action-at-a-distance entailed 
by Interpretation 2.

Summarizing, Einstein thought that any possible interpretation in
which the wave function was regarded as a complete description of 
physical reality would have to entertain unacceptable 
(relativity-violating) action-at-a-distance associated with the
collapse of the wave function on measurement. Thus he rejected 
Interpretation 2 in favor of Interpretation 1, according to which 
the wave function was an incomplete description of individual 
processes, to be (presumably) completed with something like de
Broglie's pilot-wave model. 

Just like the (later) EPR argument, this simple thought experiment 
was intended to pose a dilemma between locality and completeness: if the 
wave function represents a complete description of physical reality, then 
physical reality contains relativity-violating action-at-a-distance. Or 
equivalently, if we insist on respecting relativity's prohibition, we
must regard the quantum mechanical description of reality as
incomplete.

As is hopefully now clear, de Broglie's re-formulation of the thought
experiment in terms of literal boxes simply exaggerates the spatial
separation of two parts of the total wave function, and thus brings
out more clearly the implications of the locality assumption. But the
essential structure of the argument de Broglie presents is identical
to the original Einstein argument of 1927.

Some years after the 1927 Solvay Conference, Einstein did discuss the 
specific scenario involving a single particle divided between two 
half-boxes, although he used the boxes as merely a classical analog to 
clarify his definitions of locality and completeness. The discussion 
occurred in a 1935 letter to Schr\"odinger, written shortly after the 
publication of the EPR paper. As Arthur Fine\cite{fine3}
pointed out, in this letter Einstein disclosed that Podolsky, not
Einstein, wrote the EPR paper and that he (Einstein) was
disappointed with how it turned out: the main point had been
``smothered by the formalism.''\cite{smother} Let us see how
Einstein reworked the EPR argument in this letter, and also
examine the role played therein by the boxes.

Einstein began the letter by clarifying what he meant by
locality and completeness. He asked Schr\"odinger to consider a
classical particle (a ball) confined to a box. A partition is then
inserted, and the two half-boxes are carried to distant
locations where they are separately opened and their contents
examined. As in his discussion at the Solvay Conference, Einstein
surveyed two possible ways of analyzing this experiment:
\begin{quote}
``Now I describe a state of affairs as follows: The
probability is 
$1/2$ 
that the ball is in the first box. Is this a complete description? \\
NO:  A complete description is: the ball
is (or is not) in the first box. That is how the
characterization of the state of affairs must appear in a complete
description. \\
YES: Before I open them, the ball is by no means in one of 
the two boxes. Being in a definite box only comes about when I lift the 
covers. This is what brings about the statistical character of the 
world of experience, or its empirical lawfulness. Before lifting the 
covers the state [of the distant box] is completely characterized 
by the number $1/2$, whose significance as statistical findings, 
to be sure, is only attested to when carrying out observations. Statistics 
only arise because observation involves insufficiently known factors foreign 
to the system being described.''\cite{fine4} 
\end{quote}

Note that the NO and YES alternatives map exactly onto 
Interpretation 1 and Interpretation 2 from the 1927
discussion. According to the NO view (and Interpretation 1),
the description of the state of the system in terms of
probabilities is incomplete, there being, in reality, an actual
fact of the matter about the location (trajectory) of the
particle. According to the YES view (and Interpretation 2),
the description in terms of probabilities 
is complete because the actual fact of the matter regarding
the location of the ball (particle) only comes into existence with
the act of measurement.

For the case of a classical particle like the ball, the YES 
view is not very plausible, and Einstein asserts that ``the man on
the street would only take the first \ldots interpretation
seriously.''\cite{ibid2}
But for a single electron, the YES
view is essentially the completeness claim of the Copenhagen
interpretation of quantum mechanics. Einstein didn't 
accept this view and wanted to argue that the first view (NO) was
the correct one not only for the classical particle but for the
electron as well. But he recognized that merely arguing from the
classical analogy wasn't adequate, and went on in the letter to
Schr\"odinger (and in several other places\cite{complete} over the
subsequent decades) to construct a complete argument. 

The necessary additional premise --- one that was taken for granted
for the case of the ball but needed to be made explicit to make the
argument rigorous for the electron --- was the
locality/separability assumption. Howard\cite{howard2}
translates Einstein, again from the same 1935 letter: ``\emph{My} way of 
thinking is now this: properly considered, one cannot [refute the 
completeness doctrine, Interpretation 2] if one does not make use of a 
supplementary principle: the `separation principle.' That is to
say: `the second box, along with everything having to do with its
contents is independent of what happens with regard to the first
box (separated partial systems).' If one adheres to the separation
principle, then one thereby excludes the second \ldots point of
view, and only the [first] point of view remains, according to
which the above state description is an 
\emph{incomplete} description of reality, or of the real states.''

Einstein continues: ``The preceding analogy [that is, the
ball-in-boxes scenario] corresponds only very imperfectly to the
quantum mechanical example in the [EPR] paper. It is, however,
designed to make clear the point of view that is essential to me.
In quantum mechanics one describes a real state of affairs of a 
system by means of a normed function $\psi$ of the coordinates (of 
configuration space). The temporal evolution is uniquely determined
by the Schr\"odinger equation. One would now very much like to say
the following: $\psi$ stands in a one-to-one correspondence with
the real state of the real system. The statistical character of
measurement outcomes is exclusively due to the measuring
apparatus, or the process of measurement. If this works, I talk
about a complete description of reality by the theory. However, if
such an interpretation doesn't work out, then I call the
theoretical description `incomplete'.''\cite{fine71}

Note that Einstein defines ``completeness'' here as a one-to-one 
correspondence between quantum wave functions and real states of
real systems. This definition is somewhat different from the one (cited 
in Sec.~\ref{sec:eprdebroglie}) used in the EPR paper; Arthur Fine refers 
to it as ``bijective completeness'' in contrast to ``EPR 
completeness.''\cite{finechap5}

Einstein then dropped the ball-in-boxes example and returned to
the original EPR scenario involving two entangled particles. He 
reconstructed the EPR logic as a simple argument from the 
separation principle to a failure of bijective completeness, modeled
after the discussion of the boxes. The argument is the following:
consider the entangled state of the two particles discussed in
EPR. Before the measurement on particle 1, the entangled wave
function attributes to particle 2 no definite position and no
definite momentum. After a measurement on 1 (whatever is
measured), the wave function collapses and particle 2 is in a
state of some definite position or some definite momentum or
perhaps some other quantity altogether -- which state of course 
depends on what kind of measurement is made on particle 1 and what
its outcome is. But that aspect (which to Bohr seemed
so central to the EPR argument) is now completely irrelevant. All
that matters is that the wave function associated with the second
particle has changed in a situation where (by the separation
principle) the actual
physical state of particle 2 has not changed. That it happens to
change to a state function that can (according to the standard
eigenstate-eigenvalue link) be interpreted as attributing a definite
value for some property (an ``element of reality'') is simply
irrelevant.\cite{ewrote}

According to Einstein ``Now what is essential is exclusively that 
[the wave functions relating to the distant particle] are in general 
different from one another. I assert that this difference is incompatible 
with the hypothesis that the $\psi$ description is correlated one-to-one 
with the physical reality (the real state). After the collision, the real 
state of [the two particle system] consists precisely of the real state of 
[particle 1] and the real state of [particle 2], which two states have 
nothing to do with one another. \emph{The real state of [particle 2] 
thus cannot depend upon the kind of measurement I carry out on [particle 1].} 
(`Separation hypothesis' from above.) But then for the same state of 
[particle 2] there are two (in general arbitrarily many) equally 
justified $\psi_{[2]}$, which contradicts the hypothesis of a one-to-one 
or complete description of the real states.''\cite{howard180}

In summary, by the separation principle, the real, physical
state of particle 2 is unaffected by the measurement at 1. Yet
according to the collapse postulate, the wave function
associated with particle 2 changes. At most one of the two wave
functions can constitute a complete description. Thus the quantum
mechanical description is shown to be, in the general case,
incomplete.

Although Einstein dropped the ball-in-boxes thought experiment (and 
returned to the two-particle entangled state of EPR) when presenting
this reformulated argument from locality to incompleteness, he need
not have done so. Despite his statement that the ball-in-boxes
analogy ``corresponds only very imperfectly to the quantum mechanical
example in the [EPR] paper,'' the EPR dilemma can be established
equally well with a quantum mechanical version of the boxes.

Consider the quantum state of the single particle whose wave
function has been split between two spatial locations (boxes): 
\begin{equation}
\psi = \frac{1}{\sqrt{2}} (\psi_1 + \psi_2),
\end{equation}
where $\psi_{1,2}$ are completely localized in two well-separated regions 
of space ($B_{1,2}$). According to the standard recipe, if and when a 
position measurement is made on the particle, its wave function immediately 
collapses to zero at every point except the place where the particle was 
found. Applying this recipe here, imagine that we open box $B_1$ and
find the particle there.
At this moment, the wave function in $B_2$ immediately goes to
zero (indicating that there is no longer any probability of
finding the particle there). But according to the locality
premise, the actual physical state of $B_2$ (whatever that might
have been) cannot have changed. Hence, we have two different
quantum mechanical descriptions (in particular, wave functions
with different values in $B_2$) of the same one unchanged physical
contents of $B_2$ -- a clear failure of Einstein's (bijective)
completeness criterion.

This reformulation of Einstein's argument for incompleteness using the 
ball-in-boxes scenario is astonishingly simple compared to the EPR argument
for the same conclusion. Our reformulation highlights that, unlike our 
reconstruction of de Broglie's boxes argument and unlike the argument 
presented in EPR, the argument from locality to 
incompleteness need not rely on actually determining the identity of
any elements of reality (by means of the EPR criterion of
reality).  In the EPR scenario involving
the two entangled particles, there is an almost irresistible
tendency to attempt to establish definite properties for the distant
particle; after all, there is a particle there that one is tempted to
think has those properties. With the (one particle) Einstein's Boxes
argument, however, the contents of the distant box are less clear,
and the temptation to attribute definite properties is consequently
reduced -- which is a good thing, from the point of view of emphasizing
that it does not matter to the argument what the contents of the box
are. All that matters is that according to the locality principle,
whatever is there doesn't change as a result of the measurement 
performed in $B_1$. That alone is sufficient to establish
the conclusion that Einstein's bijective completeness criterion
fails for (an assumed local) quantum theory.

It also bears repeating that Einstein's actual argument in terms 
of the pair of previously interacting particles and the reformulation 
here in terms of the ball-in-boxes scenario both completely avoid any 
mention of the non-commuting observables that seemed (to Bohr at least) 
to play such a large role in the actual EPR paper. It is intriguing 
to speculate about how Bohr would have responded to this simplified
version of the argument for incompleteness.  We won't, however,
undertake such
speculation here.  Suffice it to credit Einstein's Boxes for helping to 
un-smother Einstein's true arguments
against the quantum completeness doctrine.

Although Einstein strongly believed in the locality/separability 
premise of this argument and 
therefore regarded it as an argument for incompleteness, he realized 
that, strictly speaking, what the argument establishes is only that, for
quantum mechanics, locality entails incompleteness. That is, 
quantum mechanics must give up either the completeness claim or the 
locality principle: 
\begin{quote}
``By this way of looking at the matter it becomes 
evident that the paradox forces us to relinquish one of the following 
two assertions:
\begin{enumerate}
\item the description by means of the $\psi$-function is complete.
\item the real states of spatially separated objects are
independent of each other.
\end{enumerate}
On the other hand, it is possible to adhere to (2) if one regards the 
$\psi$-function as the description of a (statistical) ensemble of systems 
(and therefore relinquishes (1)). However, this view blasts the framework 
of the `orthodox quantum theory'.''\cite{ereply}
\end{quote}
Or as Arthur Fine eloquently summarizes this point: ``It 
is important to notice that the conclusion Einstein draws from EPR is not a 
categorical claim for the incompleteness of quantum theory. It is rather 
that the theory poses a dilemma between completeness and separation; both 
cannot be true.''\cite{fine378}

Perhaps this elaboration of the logic of Einstein's argument against 
the completeness doctrine --- especially the simplified version
involving the boxes --- helps explain why
Einstein never believed that Bohr (or anyone else) had adequately
addressed the dilemma between locality and completeness posed in
EPR.\cite{jammer187}

\section{Bell and the Boxes}
\label{sec:bell}

In the last paper of his tragically short life, J.\ S.\
Bell prefaced a discussion and derivation of his
celebrated inequality with a classical analogy reminiscent of
Einstein's boxes. Bell wrote:
\begin{quote}
``Most physicists were (and are) unimpressed by [the correlations 
exhibited in the EPR scenario]. That is because most physicists do not 
really accept, deep down, that the wavefunction is the whole story. 
They tend to think that the analogy of the glove left at home is a good 
one. If I find that I have brought only one glove, and that it is 
right-handed, then I predict confidently that the one still at home 
will be seen to be left handed. But suppose we had been told, on good 
authority, that gloves are neither right- or left-handed when not 
looked at. Then that, by looking at one, we could predetermine the 
result of looking at the other, at some remote place, would be 
remarkable. Finding that this is so in practice, we would very soon 
invent the idea that gloves are already one thing or the other even 
when not looked at. And we would begin to doubt the authorities that 
had assured us otherwise. That common-sense position was that taken by 
Einstein, Podolsky and Rosen, in respect of correlations in quantum 
mechanics. They decided that the wavefunction, making no distinction 
whatever between one possibility and another, could not be the whole 
story. And they conjectured that a more complete story would be 
locally causal. 

``However it has turned out that quantum mechanics can not be 
`completed' into a locally causal theory, at least as long as one 
allows, as Einstein, Podolsky and Rosen did, freely operating 
experimenters. The analogy of the gloves is not a good one. Common 
sense does not work here.''\cite{bellnew}
\end{quote}

Bell goes on in the paper to derive his famous theorem, in the context 
of Bohm's version of the EPR argument.\cite{bohmbook}
In this version, it is different spin components of the two particles, 
rather than their positions and momenta, which play the role of quantum
mechanically non-commuting variables which an EPR-like argument
suggests may nevertheless possess simultaneously definite values. 

Assuming a non- (or super-) quantum-mechanical theory in which
each particle in the EPR pair carries with it a set of hidden
variables (an ``instruction set'') telling it how to respond when
encountering variously oriented polarizers (that is, a theory with
hidden elements of reality of precisely the kind suggested by the 
EPR-Bohm thought experiment), Bell imposes a locality constraint
according to which the outcome of the measurement on each side
must depend only on the variables in the particles' instruction
set and the orientation of the local apparatus. The dependence of
each outcome on the setting of the \emph{distant} apparatus or the 
\emph{distant} outcome is excluded. The mathematical consequence 
is an inequality restricting the average correlations between the 
outcomes of measurements on the two particles. 

This kind of inequality is violated by the quantum mechanical
predictions and, it seems, by real particles in real
experiments.\cite{weihs} This violation means that one (or more)
of the assumptions made in deriving the inequality must be false.
The two assumptions typically called into question are the assumption
of hidden variables and the assumed lack of (relativity-violating)
causal dependence of the outcomes on distant settings and distant
outcomes.  And choosing one of these assumptions to blame 
has generally been considered no dilemma at all; relativity's
prohibition on super-luminal causation is a central pillar of modern
physics, so ``obviously'' the correct choice is to reject the
assumption of hidden variables. Thus, the
empirical violations of Bell's inequalities seem to be regarded by
most physicists as the best argument against the hidden
variables program and for the completeness doctrine.

For example, N. David Mermin\cite{mermin}
wrote that violations of Bell's inequalities ``fatally undermine
the position of EPR'' meaning, presumably, EPR's claim that a
(hidden variable) completion of quantum theory might be possible.
Daniel Styer, answering an alleged ``misconception regarding quantum
mechanics''\cite{styer},
stated that ``The appealing view that a state vector describes an
ensemble of classical systems [and therefore an incomplete
description of individual systems] was rendered untenable by tests
of Bell's theorem which show that no deterministic model, no
matter how complicated, can give rise to all the results of quantum
mechanics.'' (Bohm's theory of course provides a straightforward
counterexample to this claim, unless Styer intends the tacit
premise that the deterministic model must respect locality.) Asher
Peres and Daniel R. Terno\cite{peres} assert that ``Bell's theorem
does not imply the existence of any non-locality in quantum
mechanics itself;'' only ``classical imitations of quantum
mechanics'' suffer inevitable non-locality. And Eugene Wigner
wrote: ``In my opinion, the most convincing argument against the
theory of hidden variables was presented by J.\ S.\
Bell.''\cite{wigner}

In summary, it is almost 
universally believed that experimental violations of 
Bell's inequalities imply fatal problems with attempts to ``complete'' 
quantum theory by adding hidden variables, but pose no particular problem 
or puzzle for quantum mechanics itself. This attitude evidently accounts 
for the fact that very few physicists 
are presently interested in the hidden variables program. It also
explains why equally few worry that there is any deep
contradiction between standard quantum theory and relativity.

Bell himself, however, did not see it this way. Like Einstein, Bell 
believed that quantum mechanics itself faced a problematic dilemma 
between completeness and locality. We will discuss the logical implications 
of this shortly, in particular, the implications of combining
the EPR (or Einstein's Boxes) locality-completeness dilemma with
Bell's theorem.

First, let us give one final version of the argument for this dilemma by 
applying Bell's reasoning to the boxes scenario. Instead of applying this 
reasoning to an EPR-motivated hidden variable theory (like Bell 
did), however, we will assume orthodox quantum mechanics, completeness 
doctrine and all, as our working theory. What we will see is that by
using reasoning similar to Bell's, we can arrive, not at a
conclusion equivalent to Bell's famous theorem, but rather at a
new argument for the EPR dilemma: quantum mechanics is either
incomplete or non-local.

Let us begin by assuming that the initial wave function,
\begin{equation}
\label{eq:3}
\lambda = \frac{1}{\sqrt{2}}(\psi_1 + \psi_2),
\end{equation}
is a complete description of the physical reality of the pre-measurement
situation.  (As before, $\psi_1$ and $\psi_2$ represent states in which the
particle is localized respectively in $B_1$ and $B_2$.
Note too that we have used the variable $\lambda$ rather than $\psi$ to
denote the pre-measurement wave function. This choice is 
meant to highlight the similarity to Bell's derivation, in which
$\lambda$ is traditionally used to denote a complete
specification of the pre-measurement state of the two particles 
including any necessary hidden variables. Here, however, we are
assuming with Bohr that the wave function alone provides a
complete specification of the pre-measurement contents of the
boxes.) 

Let $N_1 = \{0,1\}$ 
represent the number of particles found in box $B_1$ when it is
opened; similarly for $N_2$.  The Born rule implies the following 
expressions for the probabilities of finding $N_1=1$ (when/if $B_1$
is opened and examined) and of finding $N_2=1$ (when/if $B_2$ is
opened and examined):
\begin{equation}
P(N_1\!=\!1 \,|\, \lambda ) = \frac{1}{2},
\end{equation} 
and 
\begin{equation}
P(N_2\!=\!1 \,|\, \lambda ) = \frac{1}{2}.
\end{equation}

Finally, let us calculate the probability for a double
detection, that is, the probability that the particle is found
in \emph{both} of the two half-boxes when they are opened. Call this
probability 
$P(N_1\!=\!1 \; \& \; N_2\!=\!1)$. If we follow exactly Bell's
reasoning in deriving an inequality for hidden variable theories,
we can (by Bayes' formula\cite{bellnew109}) decompose the joint 
probability into a product as:
\begin{eqnarray}
P(N_1\!&=&\!1 \; \& \;N_2\!=\!1 \,|\,\lambda) \nonumber \\
 &=& P(N_1\!=\!1\,|\,\lambda)\; \cdot \; 
P(N_2\!=\!1\,|\,N_1\!=\!1,\lambda).
\end{eqnarray}
Now, quoting Bell: ``Invoking local causality, and the assumed 
completeness of \ldots\ $\lambda$ \ldots, we declare redundant
certain of the conditional variables in the last expression,
because they are at space-like separation from the result in
question.''\cite{bellnew109} In our situation, this same
local causality requirement implies that 
\begin{equation}
P(N_2\!=\!1 \,|\, N_1\!=\!1, \lambda ) = P(N_2\!=\!1 \,|\, \lambda),
\end{equation} 
because, as Bell says, the outcome event at $B_1$ ($N_1=1$) is (let
us assume) at a space-like separation from the event in question
($N_2=1$) at $B_2$. Thus, the probability of this event at $B_2$
should not depend on the outcome
$N_1=1$; it should instead depend only on those variables 
($\lambda$) that described (by assumption, completely) the
pre-measurement contents of the boxes.

Thus, the probability of double-detection simplifies to
\begin{eqnarray}
\label{jointdetectprob}
P(N_1\!&=&\!1 \, \& \, N_2\!=\!1 \,| \lambda) \\
&=& \, P(N_1\!=\!1\,|\,\lambda)\, \cdot \, P(N_2\!=\!1\,|\,\lambda).\nonumber
\end{eqnarray}
If we substitute the two 50\% probabilities on the right-hand
side, we evidently find that a double-detection should occur with
probability 25\%. That is, one quarter of the time, we should find
the particle in both boxes.

One expects, of course, that this will never happen. (Such an
occurrence would violate several fundamental
conservation principles.) Therefore one (or more) of the
assumptions leading to this prediction must be wrong. In
particular, it seems that we must reject either the assumption of
locality or the assumption of completeness -- the same dilemma
established earlier using the boxes in a different way.\cite{finepc}

The reader might object that we neglected to collapse
the wave function and thus erred in treating the second
measurement as based on the same quantum state ($\lambda$) as the
first. As Einstein pointed out in 1927,\cite{jam6}, it is precisely 
the collapse postulate which
prevents a single particle, whose wave function before measurement
is spread out over some finite region of space, from being detected
at two different places at once.

Our claim, however, is that this collapse violates the locality assumption 
if the wave function description is regarded as complete. If the 
physical contents of the second box do not change as a result of the 
measurement on the first box, and if these physical contents are 
completely described by the wave function $\lambda$, then the 
associated (Born rule) probability cannot change as a result of the outcome 
of the distant experiment. That is precisely Bell's locality assumption. 
To demand the use of a new, collapsed wave function for 
the second box (one that takes into account the information gathered when 
the first box was opened) is to concede that the description provided by 
the initial wave function was incomplete. 

If the reader finds this response unconvincing, imagine that the
same objection was leveled against the parallel assumption in 
Bell's theorem by an advocate of a supposedly local hidden
variable theory. Such an advocate might object to Bell on the
grounds that, after the first measurement is made (on particle $A$
say), the state specification of particle $B$ should be allowed to
re-adjust to be in accord with the outcome of (that is, the 
information obtained during) the experiment on $A$. ``The
experimental violation of Bell's inequalities,'' such a person
might say, ``in no way proves that my local hidden variable theory
is untenable. For Bell unjustifiably assumed that the
probabilities for various outcomes at $B$ could depend only on the
physical state attributed to $B$ before the measurement at $A$. But
in my local hidden variable theory, the state attributed to $B$,
and therefore the probability for a particular outcome at $B$,
depends on the result of the measurement at $A$.  Thus Bell's 
constraint does not apply and my theory is saved.''

The correct response to such an argument would be to simply deny that 
this sort of theory is local. A state attribution for $B$ that
changes as a result of the measurement outcome at $A$ is precisely
the kind of non-locality Bell's locality assumption is meant to
exclude. But this response applies equally well 
to quantum theory itself, if we are assuming that quantum theory
provides, with the wave function alone, a complete description of
physical reality. Recalling Bell's reasoning, it is precisely the
assumption of ``the assumed completeness of \ldots\ $\lambda$''
that warrants the removal of the $A$-side outcome from the list of
variables on which the $B$-side result is conditioned. 

Despite the conventional wisdom to the contrary, Bell's
reasoning is not uniquely applicable to hidden variable theories
or any other particular kind of theory. Rather it is a way of
teasing out the implications of any theory that is assumed to be 
both complete and local.  If we use the Einstein's Boxes scenario, this 
reasoning (applied to quantum theory without any hidden variables) 
provides yet another way to establish
Einstein's dilemma: quantum theory either violates relativity's
demand for locality/separability or is incomplete.\cite{avoid}

Note that this Bell-inspired argument for the quantum completeness-locality 
dilemma relies on definitions of completeness and locality that are 
different from those encountered in Secs. \ref{sec:eprdebroglie} and 
\ref{sec:einstein}.  Locality is defined here formally in terms of
the stochastic irrelevance (or redundancy) of facts that are
outside the past light cone of a given event: 
$P(N_2\!=\!1\,|\,N_1\!=\!1,\lambda ) = P(N_2\!=\!1\,|\,\lambda)$. 
And completeness here amounts to the standard assertion that the
probability of detecting a particle is given by the absolute
square of the wave function, and that this probability cannot be
accounted for by ignorance regarding any additional hidden
variables -- i.e., that there is nothing
other than the value of $\lambda$ in $B_2$ that determines
the probability for detecting a particle there.\cite{given}

We return then to the earlier point about the use of Bell's theorem 
as an argument against the hidden variables program (that is,
its use as an argument for the orthodox completeness
doctrine). It seems that we need to reconsider the implications of
Bell's theorem when it is combined with the completeness-locality
dilemma. If Bohr can be considered to 
have adequately refuted the argument for this dilemma, that is, if 
quantum mechanics can be validly interpreted as both complete and local, 
then Bell's theorem would indeed argue
strongly against any attempt to supplement the theory with
additional variables. But what if (perhaps motivated by Einstein's Boxes) 
we regard Bohr's response
as inadequate? That is, what if we agree with Einstein that a
dilemma between locality and completeness exists for quantum
theory?

There seem to be two options. We might choose to regard the
quantum theory as complete but non-local. But then we are no longer 
in a position to use the required non-locality of hidden variable 
theories (required, that is, by Bell's theorem) as a reason to
dismiss them. If non-locality is unacceptable to physical theories,
not just Bohm's theory (for example), but quantum theory itself would
have to be dismissed as in conflict with relativity theory.

Perhaps we are therefore tempted to back up and take the other horn
of the dilemma -- to regard quantum mechanics as
local but incomplete. But then it seems clear that we ought to
attempt to complete it, presumably by adding hidden variables
(which, it turns out, will have to be non-local). 

Either way, Bell's theorem ceases to be a valid argument against the
hidden variables program -- a conclusion opposite the standard 
interpretation of Bell's result. 

If the argument for the
locality-completeness dilemma is sound -- and I think the various
formulations in terms of Einstein's Boxes make this argument
rather compelling compared to the original EPR argument -- then
Bell's result turns into a powerful argument in favor of taking
seriously non-local hidden variable theories like Bohm's.\cite{logic}

\section{Heisenberg, Locality, and Experimental Metaphysics}
\label{sec:locality}

Heisenberg also discussed something like Einstein's
Boxes: ``\ldots\ one other idealized experiment (due
to Einstein) may be considered. We imagine a photon which is
represented by a wave packet built up out of Maxwell waves. It will
thus have a certain spatial extension and also a certain range of
frequency. By reflection at a semi-transparent mirror, it is possible
to decompose it into two parts, a reflected and a transmitted packet.
There is then a definite probability for finding the photon either in
one part or in the other part of the divided wave packet. After a
sufficient time the two parts will be separated by any distance
desired; now if an experiment yields the result that the photon is,
say, in the reflected part of the packet, then the probability of
finding the photon in the other part of the packet immediately becomes
zero. The experiment at the position of the reflected packet thus
exerts a kind of action (reduction of the wave packet) at the distant
point occupied by the transmitted packet, and one sees that this
action is propagated with a velocity greater than that of
light.''\cite{heisenberg}

There are several interesting points to note. 
First, Heisenberg converted the Boxes thought experiment into a more
realistic (and experimentally realizable; see the following)
situation where the carrying apart of two half-boxes is replaced by
the natural motion of a particle after being either transmitted or
reflected by a half-silvered mirror. Second, it is interesting that
Heisenberg seemed to accept the dilemma between locality and
completeness that Einstein's argument is intended to establish and to
favor the ``complete but non-local'' response to that dilemma. 

Yet as he claimed in the sentence just following the previous quote,
the admitted ``action at a distance'' that is exerted by the 
nearby experimental apparatus doesn't lead to a contradiction with 
relativity: ``However, it is also obvious that this kind of action
can never be utilized for the transmission of signals so that it
is not in conflict with the postulates of the theory of
relativity.'' \cite{heisenberg}

So evidently Heisenberg's position was to concede to Einstein that 
quantum mechanics was non-local, but to deny that its particular type
of non-locality was at odds with relativity's prohibitions on 
superluminal causation. (This view is widely held and was eloquently
advocated recently by Abner Shimony\cite{abner} who described the
situation as a ``peaceful coexistence'' between quantum non-locality
and relativity.\cite{referee}) The tacit premise here is that the
only kind of superluminal causation that relativity theory forbids is
the kind that can be used ``for the transmission of signals.''

This particular attempt to wriggle free of the dilemma was assessed 
by Bell:
\begin{quote}
``Do we then have to fall back on `no signaling faster than light' as the 
expression of the fundamental causal structure of contemporary theoretical 
physics? That is hard for me to accept. For one thing we have lost the 
idea that correlations can be explained, or at least this idea awaits 
reformulation. More importantly, the `no signaling' notion rests on concepts 
which are desperately vague, or vaguely applicable. The assertion that 
`we cannot signal faster than light' immediately provokes the question:

Who do we think \emph{we} are?

\emph{We} who can make `measurements', \emph{we} who can manipulate 
`external fields', \emph{we} who can `signal' at all, even if not 
faster than light? Do \emph{we} include chemists, or only physicists, 
plants, or only animals, pocket calculators, or only mainframe 
computers?'' \cite{bellnew111} 
\end{quote}

In addition, Heisenberg's move here undermines one of the principal 
arguments against non-local hidden variable theories like Bohm's.
For although that theory does include non-local,
action-at-a-distance effects, these effects (just like the similar
effects Heisenberg concedes are present in orthodox quantum
mechanics) cannot be used to transmit signals. In Bohm's
theory this inability arises from the inevitable uncertainty about the
initial locations of particles within their guiding waves. So if, as
Heisenberg argues, it is only the possibility of sending superluminal
signals that conflicts with relativity, we must evidently conclude
that Bohm's theory and orthodox quantum mechanics enjoy an equally
peaceful coexistence with it.\cite{maudlinclear}

Let us finally return to the previous comment about the
apparent experimental realizability of the Heisenberg version of
Einstein's Boxes. According to John Clauser,\cite{clauser}
Schr\"odinger was intrigued by the clarity with which Einstein's
Boxes brought out the locality-completeness dilemma for quantum
theory, and persuaded \'Ad\'am, J\'anossy, and Varga
(AJV) to perform Heisenberg's version of the experiment.\cite{ajv}
Let us quote from Clauser's description of the AJV experiment:
\begin{quote}
``In their experiment, two independent photo-detectors are placed 
respectively in the transmitted and reflected beams of a half-silvered 
mirror. If photons have a particle-like character, that is, if their 
detectable components are always spatially bounded and well localized, 
then photons impinging on the half-silvered mirror will not be split in 
two at this mirror. On the other hand, if they are purely wavelike in 
nature, (as with classical waves) then they can and will be split
into two independent classical wave packets at the mirror. This
fact then implies that if they are purely wave-like (in this
classical sense), then the two detectors will show coincidences
when a single temporarily localized photon is directed at said
mirror. One of these independent wave packets will be transmitted
to illuminate the first detector, and the other will be reflected
to illuminate the second detector, and both detectors will then
have a finite probability of detecting the same photon (classical
wave packet) \ldots\ [AJV] found no such anomalous coincidences,
and thereby concluded that \ldots\ photons do not split at the
mirror and thus do exhibit a particle-like
character.''\cite{clauser} 
\end{quote}

Note that this experiment is precisely an empirical test of
Eq.~(\ref{jointdetectprob}) . Although
the outcome of the experiment is not surprising to anyone
today, it is hoped that the preceding discussion underlines the
apparent implications of this outcome: if we accept the locality
principle, we must evidently conclude that
photons\cite{photons} do not split at the mirror, but instead
choose one or the other path exclusively. That is, we must
conclude that the quantum mechanical description of the photon
before detection,
\begin{equation}
\psi = \frac{1}{\sqrt{2}}(\psi_1 + \psi_2),
\end{equation}
is incomplete because it fails to attribute a definite position to
the particle. 

Alternatively, we could uphold the completeness of the wave function 
description by denying the locality premise and regarding the collapse 
of the wave function on measurement as describing a real physical change 
in the state of the distant packet's referent.

There is thus an interesting parallel between the AJV experiment
and the recent experimental tests of Bell's inequalities. \cite{weihs}
Because Bell's inequality represents a constraint on
local hidden variable theories, empirical violations of the
inequalities imply that local hidden variable theories aren't
empirically viable. Likewise, Eq.~(\ref{jointdetectprob})
represents a constraint on a local version of quantum theory (assumed
complete). So the AJV finding that this constraint is violated has
implications paralleling those of the Bell tests: the local version
of quantum theory (assumed complete) is not empirically viable.
Either locality or the orthodox completeness assumption must be
abandoned. This conclusion is not new, but the same dilemma we have
seen now several times. 

The extent to which orthodox quantum theory and the
hidden variable alternatives are on equal footing is 
striking: local versions of each are evidently refuted by
experiment.\cite{nosurprise} Thus it seems the program of
``experimental metaphysics'' (to use one of Shimony's apt
phrases\cite{abner}) began more than 30 years before the recent
tests of Bell's inequality.

\section{Discussion}\label{sec:discussion}

We have presented several different formulations and revisions of the 
argument that Einstein first presented at the 1927 Solvay
conference. The goal of this argument in all of its versions is to
establish an inconsistency between the following two claims:

\begin{itemize}

\item The quantum mechanical description of reality is complete.

\item Quantum mechanics describes a world with only local
interactions.\cite{after}

\end{itemize}
Although the various formulations of the argument differ slightly in terms
of logical structure and definitions of key terms, they point to the
same conclusion and ultimately pivot around the same fulcrum: 
the collapse of the wave function.

The collapse rule itself is absolutely essential to the quantum formalism,
as evidenced, for example, by the derivation of
Eq.~(\ref{jointdetectprob}) and the subsequent discussion in
Sec.~\ref{sec:bell}.
However, the exact meaning of the rule is somewhat opaque in the standard 
interpretation. The essential lesson of Einstein's Boxes (and 
this is a point Einstein seems to have understood quite clearly as early 
as 1927) is that we face a very simple choice when it comes to interpreting
the collapse. Either it represents a real, physical change of the 
physical state of the affected system, or it
doesn't.\cite{similar} If it does represent such a physical
disturbance, this entails a non-local
action-at-a-distance that seems to conflict with relativity's
prohibition on superluminal causation. (Or, if the physical
disturbance produced by the collapse were to propagate from
the measurement event at a sub-luminal speed, the quantum
predictions would be empirically wrong in cases of spatially
separated correlations like those illustrated in the AJV
experiment.)  On the other hand, if the collapse of the wave
function does not represent a physical change in the state of the
real system, then the quantum mechanical description
must be incomplete (in at least Einstein's bijective sense). For
quantum theory would then attribute two distinct state 
descriptions to one and the same physical state.

The orthodox claim, initiated by Bohr in his reply to EPR, is that
quantum mechanics is both complete and local. On further scrutiny, 
however, it seems this claim can be maintained only with a kind of 
double-speak about
the collapse postulate. To protect the completeness claim, the
collapse must be interpreted as a real physical state change,
usually backed up by some version of the disturbance theory of
measurement.\cite{brown} On the other hand, the locality claim can
be defended only by interpreting the wave function collapse as
merely a change in knowledge.\cite{fuchs}

This double-speak was perhaps first identified by Einstein. In the 1935 
letter to Schr\"odinger that was discussed earlier, Einstein wrote:
``The talmudic philosopher doesn't give a hoot for `reality', which he 
regards as a hobgoblin of the naive, and he declares that the two points 
of view differ only as to their mode of
expression.''\cite{howard178}
The ``Talmudist'' here is Bohr, and the ``two points of view'' are the
two interpretations of the boxes discussed previously. But these 
interpretations map precisely onto the two just-mentioned
interpretations of wave function collapse. If, on the one hand, the
wave function description is incomplete (Interpretation 1 from 
Sec.~\ref{sec:einstein}), then the
collapse postulate can be safely interpreted as merely updating our
knowledge. But if, on the other hand, the wave function description is
regarded as complete (Interpretation 2), then the collapse must
represent a real physical change in the state of the system.

Evidently Einstein believed that the claim that wave function collapse
represents a physically real change of state, and the claim that it
represents a mere change in knowledge, do not ``differ only as to
their mode of expression.'' 

Of course, the existence of a clean distinction between these two
views of wave function collapse presupposes a realist philosophy
that Bohr and many of his followers deny (or even claim to have
refuted). Bohr, for example, is said to have
declared that ``there is no quantum world.''\cite{mermintoday}
Although this anti-realist attitude continues to receive lip service,
it seems that most physicists don't take the literal
denial of a quantum world very seriously.\cite{serious} Most of us
believe in the real existence of atoms, electrons, and other
sub-atomic particles. Whatever strange quantum properties they may
have, surely there is a ``they'' that have them. 

We will by no means resolve the debate about realism here. But it is 
important to recognize that anti-realism is not a sensible 
response to the alleged incompatibility of the completeness and 
locality claims. It is, rather, a wholesale denial of the very
issues of completeness and locality: there is no sense, for
example, to a claim that there is no quantum world, but quantum
mechanics provides a complete description of it.\cite{stapp}

One cannot respond to Einstein's dilemma by claiming that
an anti-realist attitude warrants a simultaneous belief in the
completeness and locality of quantum mechanics. For according to
the anti-realist attitude, there is simply no such issue as
completeness (there being no quantum world for quantum mechanics
to completely or incompletely describe) and likewise no such
issue as locality. This wholesale denial of the issues is not what
Bohr seems to have expressed explicitly, and it does not seem
consistent with the views of most contemporary defenders of the
Copenhagen interpretation who think, for example, that locality 
is a meaningful and important requirement for physical theories.

It is only from the point of view of the quantum realist that
these issues are meaningful; and from this point of view, they
are crucial and must be addressed before we can claim to truly 
understand the meaning and status of quantum theory. That is a program
Einstein helped initiate in the 1920s and, sadly, it is one that
has been largely abandoned by the physics community -- no doubt
because many came to believe that the standard interpretation was
adequate on both counts. But, from the 
point of view of Einstein, this adequacy was largely
overestimated: the standard interpretation of quantum theory
does suffer from a troubling dilemma, one with major
implications not only for the theory itself but also for possible 
alternatives to or extensions of it. Indeed, when combined with the
empirical violations of Bell's inequalities, Einstein's dilemma has
major implications for our understanding of nature.

It is fascinating that, after more than 75 years of
heated discussion, Einstein's criticisms of the quantum theory
still seem effective and fresh. There is still much work to be
done before we achieve total clarity about the foundations of
quantum theory, the kind of clarity that Einstein continuously
sought. As we approach the centennial celebration of Einstein's
``miraculous year'', let us acknowledge that Einstein still has much 
to teach us about how this clarity can be achieved.  It is hoped
that the Einstein's Boxes argument in particular will continue to
stimulate further discussions on these fascinating and important
topics.

\begin{acknowledgments}
Thanks to three anonymous referees and to Roderich Tumulka, Arthur
Fine, and Sheldon Goldstein for helpful discussions and critical
comments on drafts of the paper. A special thanks is also due to
Arthur Fine and Don Howard, on whose Einstein scholarship my own
understanding (and much of the present work) is largely based.
\end{acknowledgments}


\begin{thebibliography}{99}
\bibitem{jam2}This is one very plausible and influential
interpretation of Bohr's reply; see, for example,
Ref.~\onlinecite{jammer}, p. 196. But other commentators
have understood Bohr's rather obscure prose differently or not
at all.
\bibitem{avoid}To avoid possible confusion, let me just state
explicitly that this Bell-inspired derivation is not meant
to show that Einstein's Boxes is somehow equivalent to Bell's
theorem. It isn't. The former establishes that quantum mechanics
itself cannot be both complete and local. The latter establishes
that no local hidden variable theory can agree with quantum
theory's predictions. Our point here is to show that one can
generate a new version of the Einstein's Boxes argument by
borrowing some of the logical steps also used by Bell
in his derivation of his inequalities. It will be crucial to
the subsequent discussion that these two items -- the
locality-completeness dilemma and Bell's theorem -- are distinct.
\bibitem{given}Given that there are, at this point, 
several distinct definitions of ``completeness'' and ``locality''
on the table, the reader may begin to suspect that the dilemma we
keep running into is merely due to an improper definition of one of
these terms, and that the trouble could be avoided with some
clever new definitions. One such attempt will be considered in 
Sec.~\ref{sec:locality}.
For now let me simply caution the reader against
this attitude. The various definitions we have seen are not truly
distinct; they all focus on different aspects of a single, shared
concept of ``completeness'' (the idea that a complete theory
shouldn't leave anything out of its description of reality) and a 
single, shared concept of ``locality'' (that what exists/occurs
``over here'' shouldn't instantaneously affect what exists/occurs
``over there''). In my opinion, the lesson of the various
formulations we are considering is not that new definitions might
allow us to elude the dilemma, but rather that any
reasonable definitions give rise to the dilemma. (As we
will see in the next section, a sufficiently radical redefinition
in which the basic meaning is changed can ``succeed'' in removing
the dilemma. But this isn't true success; it's cheating.)
\bibitem{maudlinclear}See Ref.~\onlinecite{maudlin} for a clear
and systematic discussion of the possible positions one might take
on this issue of the consistency of quantum non-locality and
relativity.
\bibitem{photons}Here, by ``photons,'' we mean, strictly speaking,
the particles that are actually detected. There is nothing here to
rule out the possibility that, in addition, there exists
(say) a wave that splits in half at the mirror. This is the story 
told by Bohm's theory: the wave packet splits in half at the mirror
and the photon particle ends up in one or the other of the two
packets.
\bibitem{nosurprise}This is not surprising from the point of view 
outlined in Ref.~\onlinecite{given}. Because
nature is non-local, it stands to reason that all local
theories would be inconsistent with experiment.
\bibitem{similar}A similar point is made in Chap. 5 of
Ref.~\onlinecite{maudlin}.
\end{thebibliography}
\end{document}